\documentclass[aps,prl,twocolumn,showpacs,amsmath]{revtex4}
\usepackage{epsfig}

\newcommand{\hm}{\mbox{H$^-$}}
\newcommand{\kpar}{\mbox{$k_{\mbox{\scriptsize par}}$}}

\newcommand{\phii}{\mbox{$\phi_{\mbox{\scriptsize ion}}$}}
\newcommand{\phif}{\mbox{$\Phi(t)$}}
\begin{document}
\preprint{}
\title{
Evidence for parallel confinement in resonant charge transfer
of \hm\,near metal surfaces
}
\author{Himadri S. Chakraborty}
\author{Thomas Niederhausen}
\author{Uwe Thumm}
\affiliation{
James R. Macdonald Laboratory, Department of Physics, 
Kansas State University, Manhattan, Kansas 66506-2604, USA
}
\date{\today}

\begin{abstract}
Using a wave packet propagation approach, we find that the 
resonant charge transfer process of \hm\,near 
a Cu(111) surface is strongly influenced by transient hybrid states.
These states originate from an ion-induced confinement parallel to
the surface together with the surface-localization character of the metal 
potential along the surface normal. 
The lowest members of these states have lifetimes of the order of 
interaction times in typical particle-surface scattering experiments.
The propagation of the electron probability
density provides clear evidence for this effect in visualizing the evolution
and the decay of these transient states.
\end{abstract}
\maketitle
%PACs descriptions:
%79.20.Rf: Atomic, molecular, and ion beam impact and interactions 
%          with surfaces 
%34.70.+e: Charge transfer
%73.20.At: Surface states, band structure, electron density of states

The investigation of electron transfer and orbital 
hybridization processes during the interaction of a projectile 
atom or ion with a metal surface is of both fundamental and practical
importance. The ensuing knowledge finds valuable use in various applied fields 
of physics, such as, 
development of ion sources, control of ion-wall interactions
in fusion plasma, surface chemistry and analysis, secondary ion mass 
spectroscopy, and reactive ion etching\cite{gauy96,shao94}.
Of basic interest is the detailed understanding of single-electron
transfer leading to either ionization or neutralization of a 
surface-scattered projectile. This process of resonant charge transfer
(RCT) has been addressed by employing different non-perturbative
theoretical methods, including single-center basis-set-expansion\cite{bahr99},
complex coordinates rotation\cite{nord92}, 
two-center expansion\cite{thumm02-98},
multi-center expansion techniques\cite{mart96}, and the direct
numerical integration of the effective single-electron Schr\"{o}dinger
equation by Crank-Nicholson wave packet propagation (CNP)\cite{bori98,
bori99,press93,thumm02}. 

Of all these methods, CNP is most flexible in the sense that
it can readily be applied to any parametrized effective potential
that may be used to represent the electronic structure of substrate
and projectile. In contrast to expansion methods that usually simplify
the target to a free-electron (jellium) metal, CNP allows for 
a significantly more detailed representation of the substrate
electronic structure, including the effect of band gaps\cite{bori98,guil99}, 
surface states\cite{bori99}, and image
states on the RCT dynamics.

The Cu(111) surface is of particular interest since (a) the affinity level
of \hm\,lies within the $L$-band gap of the surface and (b) it 
serves as a prototype
of a metal surface that can localize a surface state within its
band gap. We show that for the \hm/Cu(111) system charge transfer
is to a large extent channeled through transient hybrid states that are
confined parallel to the surface by the combined influence of
surface and projectile potentials.
\hm\,is described by an effective potential 
that models the interaction of the active electron with a polarizable
core\cite{cohe86}. 
A one-dimensional
[in the co-ordinate ($z$) of surface normal] single-electron 
effective potential, 
constructed from pseudopotential local density
calculations, is employed to model the surface\cite{chul99}.
This potential reproduces the observed and/or {\em ab\,initio} 
$L$-band gap position, surface state and image states for zero
electron momentum component \kpar\, parallel to the surface ($x$ direction).
Note that the first image state lies in the band gap while higher ones
are degenerate with the conduction band\cite{chul99}.
We employ the CNP\cite{press93,thumm02} of
the initial free \hm\,wave function \phii
over a two-dimensional numerical grid in which the metal continuum 
is approximated by free electronic motion in $x$ direction. 
Our grid includes 100 layers on the bulk and extends to $z$ = 
200 a.u.\ on the vacuum side. The 
topmost layer of lattice points defines $z$ = 0. The grid covers 200 a.u.\ in $x$.
The grid spacings $\Delta z$ = $\Delta x$ = 0.2 a.u.\  yield  
good convergence.
\begin{figure}
\vskip -1.1cm
\centerline{\psfig{figure=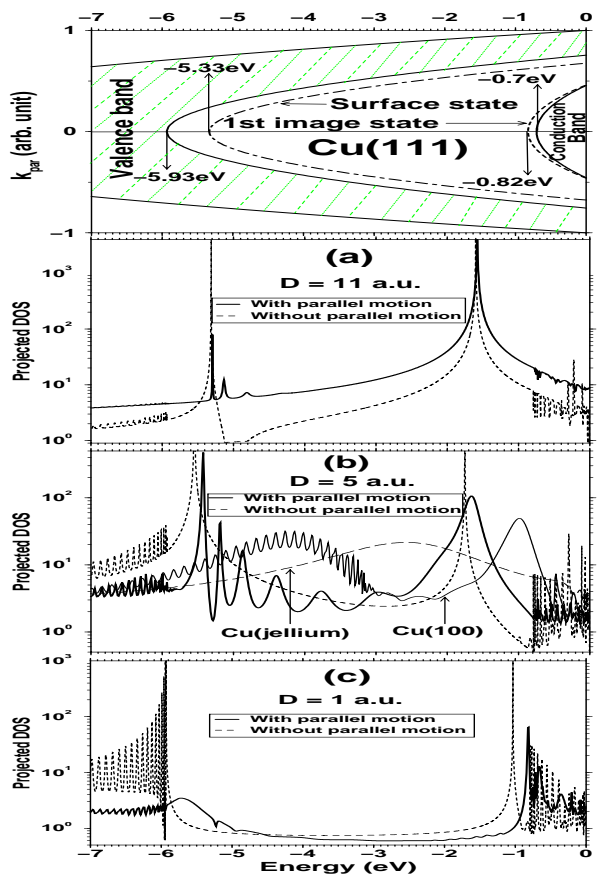,height=14.0cm,width=10.5cm,angle=0}}
\vskip -1.3cm
\caption{
Top panel: Schematic of the Cu(111) band structure as a function of
\kpar~ showing a surface state at $-$5.33 eV 
within the gap separating the valence
and the conduction band; (a) PDOS for $D$ = 11 a.u.\ for Cu(111);
(b) $D$ = 5 a.u.\ for Cu(111), Cu(100), and Cu-jellium; (c) $D$ = 1 a.u.\ 
for Cu(111).
}
\vskip -0.7cm
\end{figure}

For fixed ion-surface distances $D$, the numerical propagation over time $t$
yields \phif\,and the ionic survival
amplitude $A(t)=\langle\phif|\phii\rangle$.
The real part of the Fourier transform (FT) of this amplitude yields 
the projected density of states (PDOS) that exhibits resonance structures.
The position, width, and amplitude of these resonances provide, respectively,
the energy, lifetime, and population of the states.
Contrary to the parametric fitting adopted
in Ref.\,\onlinecite{bori99}, a direct FT
of $A(t)$ is performed by propagating, in time-steps 
$\Delta t$ = 0.1 a.u., over a period long enough
for acceptable convergence.
Figure 1 depicts the PDOS (thick solid curve) for three typical 
values of $D$. Results neglecting the electronic motion 
parallel to the surface are obtained by propagating over a one-dimensional 
grid along $z$ and are shown (short-dashed curve) for
comparisons. Note, although for this 1-D propagation, due to the absence 
of any decay continuum, $A(t)$ never fully converges, 
we still carry out the FT of 
$A(t)$ calculated
over a finite time, since we are 
interested only in identifying resonances 
in the 1-D PDOS. 

At $D$ = 11 a.u.\ [Fig\,1(a)], our results with or
without the parallel motion included show discretized structures corresponding
to the valence and conduction band. The affinity level resonance 
(at $-$1.56 eV) and
the surface state resonance (at $-$5.31 eV) are also present in both 
calculations, although the affinity level is shifted downward
from the unperturbed asymptotic affinity of $-$0.76 eV. 
Strikingly,
two small peaks appear just above the surface state resonance
for the results that include the parallel motion. 
For $D$ = 5 a.u.\ [Fig.\,1(b)], the affinity level and
the surface state resonance are present in both results, with and
without parallel motion, {\em but} the structures in between 
for the calculation that incorporates electronic parallel motion increase
in number and strength. Clearly, these new resonances 
appear only when the electronic parallel degree of freedom
is switched on. 
For a jellium Cu surface the PDOS [Fig.\,1(b), long-dashed curve] 
only shows a wide affinity level peak, as expected, thus indicating
that the resonances are due to details in the surface band structure
that are not accounted for in the simplistic jellium model.
For a more complete portrayal
of the origin of these features we
also show the PDOS [Fig.\,1(b), thin solid curve], 
including the parallel motion, for Cu(100),
which has a similar band gap in the direction normal to the surface
but, contrary to Cu(111), {\em no} surface 
state inside the gap\cite{chul99}. 
This shows valence band structures up to
$-$3.1 eV and the affinity level at $-$1 eV  
{\em but} no significant feature in between.  
Evidently, the extra features in the Cu(111) PDOS
between the surface state and the affinity level resonances
{\em must} be originating both from the parallel 
degree of freedom of the electron
and the special localizing property of the Cu(111) potential
along the surface normal that binds a surface state inside the band gap. 
For very close ion-surface separation, $D$ = 1 a.u.,
these features almost disappear, while two similar resonances 
superimposed on the conduction band spearhead, at $-$0.67 and 
$-$0.35 eV [Fig.\,1(c)], above the affinity level. 
These resonances, in analogy with the ones below the affinity level,
also originate from the electron parallel motion and the weak binding
of the long-range tail of the surface potential.

The origin and evolution of this effect with decreasing $D$ can be 
understood as follows. As \hm\,moves towards 
the surface, the ion potential gradually deepens following, at large $D$, 
the classical image 
interaction. Consequently, the spherical symmetry
of the ion potential gets broken by the surface potential
``slope'', which becomes steepest in the vicinity of metal-vacuum interface. 
Sufficiently close to the surface, the 
parallelly-stretched asymmetric top of the ion potential 
confines a new state. Although
this state is  bound in the parallel direction, 
whether or not it will
live long will depend on how much binding it
experiences in the direction normal to the surface. Indeed, for
Cu(111) the surface potential has enough reflectivity to enable the
formation of a localized surface state within the band gap. 
As a consequence, the new state, confined parallelly by the incoming
ion, is also (temporarily) bound in the normal direction by the 
Cu(111) potential.
This state is relatively long-lived and appears as a fairly narrow peak
in the PDOS spectrum [Fig\,1(a)]. As the ion moves closer 
to the surface, the number of states confined parallelly 
increases and additional peaks emerge in the PDOS [Fig.\,1(b)].
For the \hm/Cu(100) system, in 
contrast, since the surface potential lacks   
sufficient surface-localizing reflectivity (it fails to support a surface
state within the band gap), the states, confined parallelly 
by the ion, decay rapidly into the metal valence band, 
as indicated by the broad bump in the valence band of Cu(100) [Fig.\,1(b)].
Therefore, these new resonances near Cu(111) are parallelly confined hybrids. 
\begin{figure}
\centerline{\psfig{figure=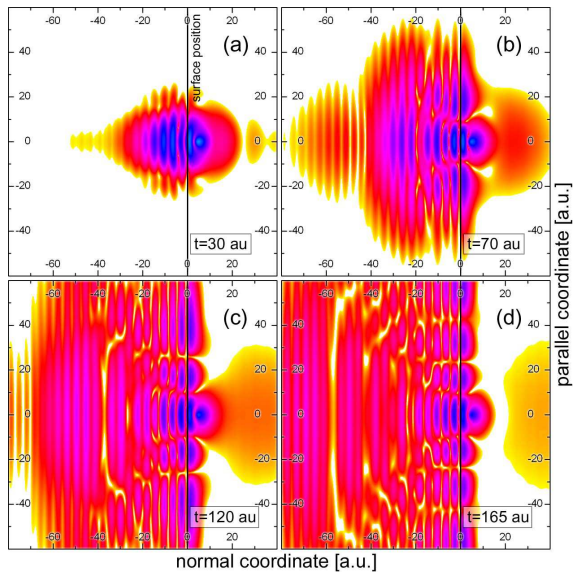,height=07.5cm,width=7.5cm,angle=0}}
\vskip 0.0cm
\caption{(Color online)
Wave packet densities (logarithmic scale)
at $t$ = 30, 70, 120, and 165 a.u.\
for propagation at a fixed ion-surface
separation of 5 a.u. The surface position is indicated by a vertical line.
}
\vskip -0.5cm
\end{figure}

The surface state is not bound in the parallel direction.
A resonance state that energetically lies above the surface state
while being confined parallelly has to be less bound in the normal
direction than the surface state.
A reduction of the normal binding can be achieved 
by sliding up the potential at the bulk-vacuum
interface, that is, by
moving the mean position of the wave function in normal direction 
towards the ion. This increases 
the overlap between the ionic wave function and that of the 
confined state. Consequently,
for a given ion-surface separation, we expect the lowest parallelly 
confined state to be populated first because of its strongest 
wave function overlap with the ion. 
This is seen in Fig.\,2, which
depicts the propagated wave packet probability 
density $|\phif|^2$ at $t$ = 30, 70, 120 and
165 a.u.\ 
for fixed $D$ = 5 a.u. Figure 2(a) ($t$ = 30 a.u.) shows
an approximately nodeless structure  
outside the surface plane
implying that over a short propagation time the ion populates
predominantly the lowest parallelly confined state. However,
at later times and increasing population of higher parallelly confined
states, the wave packet spreads along the 
parallel co-ordinate forming additional nodal structures. 
Notably, since
the parallel component of the wave packet outside the surface
is a time-dependent linear combination 
of parallelly confined stationary states 
with {\em different} nodal structures, the position
of a node moves with time.
The ripples seen along 
the normal direction inside the bulk are due to the 
periodic bulk potential
and a faint blob beyond $z \approx$ 20 a.u.\ on the vacuum 
side [Fig.\,2(b-d)] represents
the evolution of weakly populated image states.
\begin{figure}
\vskip -1.1cm
\centerline{\psfig{figure=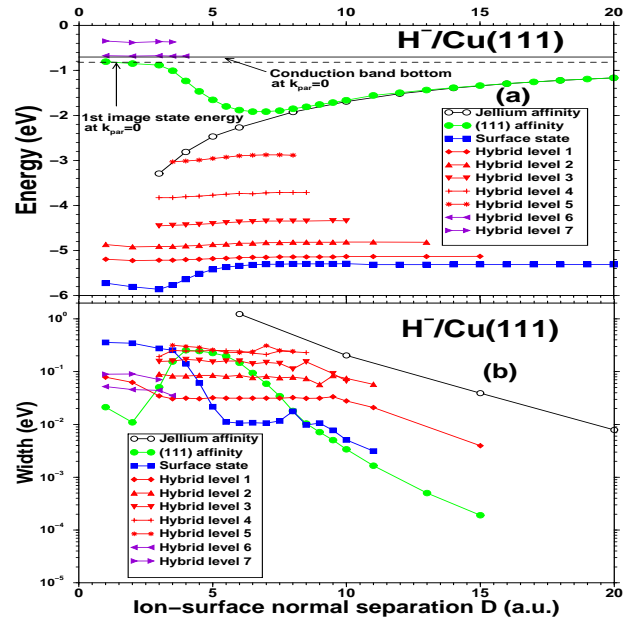,height=11.0cm,width=10.0cm,angle=0}}
\vskip -2.0cm
\caption{(Color online)
Energies and widths of various resonances as a function of the ion-surface
distance.
}
\vskip -0.5cm
\end{figure}

Fig.\,3 shows the energy and the width of various resonances as a function
of $D$. Our results for affinity level
and surface state resonances are qualitatively
similar to previous calculations\cite{bori99}. 
At large distances, the energy [Fig.\,3(a)] of the affinity level 
resonance (filled circles) 
is solely governed by image interactions leading to a good
agreement with corresponding jellium results (opaque circles). In 
Fig.\,3(b) on the other hand, affinity level resonance widths 
for Cu(111) are smaller than
the jellium predictions at large distances, since in the jellium case 
no band gap exists and electrons can decay in the normal direction.
The strong interaction, seen in the distance-dependent widths
and energies between the affinity level resonance and the surface state 
resonance (filled squares)
for small $D$, is the
consequence of an indirect coupling between the corresponding discrete
quasi-stationary
states through the surface state continuum\cite{bori99}.
Interestingly, both energy and width of the parallelly confined
resonances depend only weakly on $D$ (Fig.\,3). We explain this 
near-stabilization by couplings of a given parallelly confined
state with both affinity and surface state that have comparable 
strength. These couplings result in opposite
level shifts and comparable rates (widths) for transitions
between the affinity and parallelly confined state and between
the parallelly confined state and the surface state.
The same argument explains the stabilization
of resonances just above the ionic resonance for very close $D$ where
the states are interacting with the ion and the conduction band.
Furthermore, as discussed before, the state with maximum binding in the 
parallel direction has the strongest overlap with the affinity level 
and the weakest with the surface state. 
As a result, while it is ``fed''
by the ion the most, it decays through the surface state continuum 
the least acquiring a narrow width [Fig.\,3(b)]. 
The counter-argument explains the large widths for
minimally confined states in the parallel direction.
\begin{figure}
\centerline{\psfig{figure=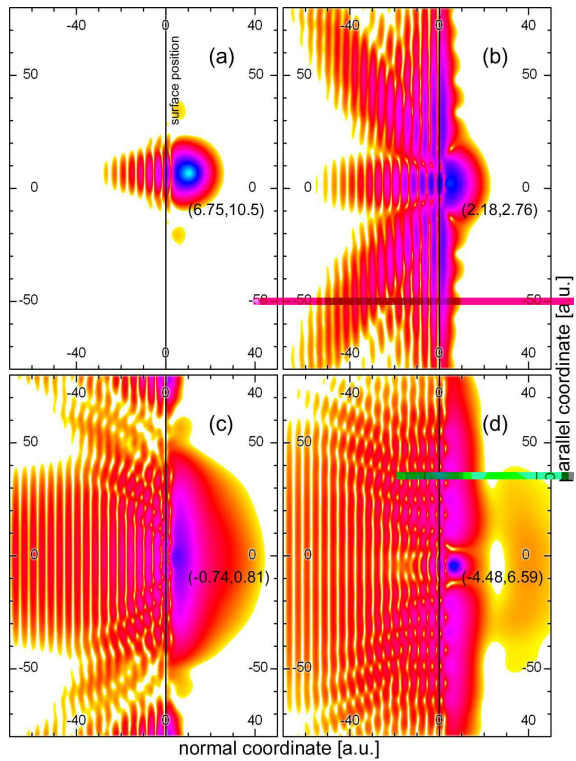,height=10.0cm,width=7.5cm,angle=0}}
\vskip 0.0cm
\caption{
(Color online) Wave packet densities (logarithmic scale)
at times $-$320 a.u.(a), $-$110 a.u.(b), 20 a.u.(c), and
180 a.u.(d), relative to the time at which the point of closest
approach is reached. The ion approaches the surface 
at an angle of 
60$^o$ with respect to the surface and with an energy of 50 eV. Positions
($X$,$D$) are given in parenthesis, with $X$ being relative to the
point of closest approach.
}
\vskip -0.7cm
\end{figure}

During the approach to the surface the projectile gradually decelerates
in the normal direction along its incoming trajectory,
owing to the repulsive interaction
between its neutral core and surface atoms,
until its normal velocity becomes zero at the point of closest approach.
For specular reflection, it 
re-gains its original normal velocity. For a given initial kinetic
energy and angle of incidence,
we simulate the classical ion-trajectory by modeling the core-surface
interaction via a plane-averaged interatomic potential\cite{bier82}.
This defines a distance of closest 
approach as a function of the initial
normal velocity.
Since the ion moves slowly near the surface,
the adiabatic (fixed-ion) results (Figs.\,1-3) provide a 
good guideline to understand 
the calculations for a moving ion. 
In Fig.\,4, we present four wave packet probability densities, 
a pair each from 
the incoming and the outgoing part of the trajectory 
of \hm\,ions with
50 eV asymptotic energy at 60$^o$ incidence
with respect to the surface.  
In Fig.\,4(a), $D$ = 10.5 a.u., the ion
predominantly populates the state confined most strongly
in the parallel direction. Reaching $D$ = 2.76 a.u., Fig.\,4(b), 
the wave packet spreads over all available parallelly
confined states, and clear nodal structures
emerge symmetrically along the parallel 
direction outside the surface with each
``bead'' emanating a jet into the bulk.
Electrons in the central jet have small parallel velocity
indicating their ejection from the most tightly confined state.
A steady increase of the parallel velocity is evidenced
going symmetrically away from the center in parallel direction 
since the more distant jets originate from less strongly
confined states.
In Fig.\,4(c), the ion arrives roughly 
at the distance
of closest approach, 0.5 a.u., 
where the adiabatic energy position 
of the ionic resonance moves very close to the conduction band [Fig.\,3(a)]
and induces new resonances above the ionic level.
Again, the node formation and resulting jets are seen, 
although the shape of 
the wave packet density is now dominated by a strong decay into the 
conduction band as well as by the subsequent population of image states,
degenerate with the conduction band,
on the vacuum side of the projectile.
A remarkable signature of the parallel confinement is finally seen 
on the outward excursion of the ion in
Fig.\,4(d) at $D$ = 6.59 a.u.: 
the entrapment of the electron back in parallelly confined states
results in decay-jets into the bulk
and subsequent re-ionization of the projectile (note the strong trapping
at the ion position).
As an observable consequence of the strong participation of 
parallelly confined states in the decay near Cu(111), we find about 6\% 
ion survival after the scattering 
of \hm\,from this surface as opposed to about 2\% from 
Cu(100), which is free from this effect. A detailed comparative
study will be published elsewhere\cite{chak03}.

In conclusion, we demonstrate
significant parallel confinement
effects in resonant neutralization of \hm\,near Cu(111) by directly
analyzing the evolution of the active electron's
wave packet probability density. A surface-induced breakdown 
of the ionic spherical symmetry and significant reflectivity
of the surface potential is responsible for this confinement. 
Finally, there is nothing special about  
Cu(111). Any of the surfaces, namely, Ag(111), Au(111), Pd(111) etc., 
supporting a surface state in the
$L$-band gap
is expected to show similar parallel confinement phenomena during the RCT 
process.

This work is supported by the NSF (grant PHY-0071035) and
the Division of Chemical Sciences,
Office of Basic Energy Sciences, Office of Energy Research, US DoE.

\end{document}